\documentclass[a4paper]{jpconf}

\usepackage{graphicx}
\usepackage{iopams} 
\usepackage{amsmath}

\usepackage{citesort}

\usepackage{epstopdf}
\usepackage[pass,a4paper]{geometry}

\begin{document}


\hfill KA-TP-31-2014

\newcommand{\referencedummy}{[...]}
\newcommand{\scriptA}{\mathcal{A}}

\title{Event simulation for colliders -- A basic overview}
\author{Christian Reuschle}
\address{Institute for Theoretical Physics, Karlsruhe Institute of Technology, Karlsruhe, Germany}
\ead{christian.reuschle@kit.edu}

\begin{abstract}
In this article we will discuss the basic calculational concepts to simulate particle physics events at high energy colliders. We will mainly focus on the physics in hadron colliders and particularly on the simulation of the perturbative parts, where we will in turn focus on the next-to-leading order QCD corrections.
\end{abstract}

\section{Introduction}

The Standard Model of Particle Physics is very successful to describe the phenomena that we observe in high energy colliders 
\cite{Beringer:1900zz}. 
In July 2012 the ATLAS and CMS experiments at CERN finally announced to have found the last missing piece of the Standard Model -- the Higgs boson 
\cite{ATLAS:2012gk,CMS:2012gu}. 
Measurements of the properties of the new boson, in order to identify it with the Standard Model Higgs boson, and searches for new physics beyond the Standard Model are ongoing since. 
In order to compare those measurements to our theoretical concepts, precise predictions for the corresponding observables are needed. However, the situation at hadron colliders is complicated through a large background of QCD radiation, and sophisticated computational techniques, theoretically as well as experimentally, are needed all the more. On the theory side Monte Carlo programs are thereby the tool of choice, in order to combine all the necessary computational techniques. In this article we will briefly describe some of the concepts that are thereby used. Due to the mixed audience, we will constrain ourselves to a rather basic description. For a more thorough description we would like to point the reader to a list of references throughout this article, which are better suited to explain certain aspects in greater detail, and we would like to apologize in advance for any relevant reference that we may miss to mention.

\section{Hadronic collisions}

\begin{figure}
\centering
\includegraphics[scale=0.95]{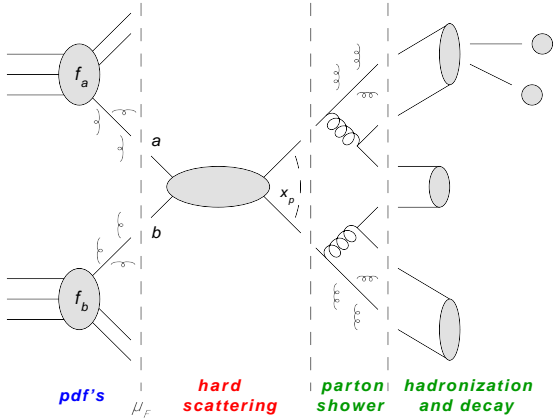}
\hspace{1.5cm}
\begin{minipage}[b]{7.0cm}
\caption{\small\it The scattering of two hadrons with PDF $f_a(x_1,\mu_F)$ and $f_b(x_2,\mu_F)$. The picture portrays a ``snapshot'' of the evolution of the hadrons during the scattering process, where we pick an arbitrary scale $\mu_F$ to factorize off the actual hard interaction, which is described by the scattering between the partonic constituents $a$ and $b$. The consecutive radiation of soft and collinear partons off the partons involved in the hard scattering is governed by a parton shower. The partons in the parton shower hadronize at a scale $\Lambda_{QCD}^2$, and may subsequently decay. Not depicted is the underlying event.}
\end{minipage}
\label{fig:lhcProcess}
\end{figure}

In this section we will sketch the physics behind general purpose event generators. Excellent general reviews on the following can be found in 
\cite{Buckley:2011ms,ellis1996:am,dissertori2003:am}, 
to which we would like to point the reader for a more thorough description and further references if required.

The current theoretical picture of a hadronic collision is motivated by the properties of asymptotic freedom of QCD at high energy scales and confinement at low scales. 
Due to the property of asymptotic freedom we may take on the parton picture, 
i.e. we assume that we can resolve gluons and quarks inside the hadrons, with a resolution scale $\mu_F$, as depicted in 
Fig.1. 
Correspondingly, the total cross section for the scattering of two hadrons at a center-of-mass energy (CME) $\sqrt{s}$ can be factorized according to the following formula: 
\begin{equation}
\sigma(s) = \sum\limits_{X_{\text{h}}} \! \sum\limits_{a,b} \! \int \!\! dx_1dx_2 \; f_a(x_1;\mu_F) f_b(x_2;\mu_F) 
\; \hat{\sigma}_{ab \rightarrow X_{\text{p}}}\!(\hat{s};\{a,b,X_{\text{p}}\};\mu_F) \otimes PS_{X_{\text{p}}} \otimes H_{X_{\text{p}}\rightarrow X_{\text{h}}}
\label{eq:lhccrosssection}
\end{equation}

Eq.\ref{eq:lhccrosssection} consists of a non-perturbative and a perturbative part. In the non-perturbative part the parton density functions (PDF) $f_a(x_1;\mu_F)$/$f_b(x_2;\mu_F)$ describe the probabilities at a scale $\mu_F$ to ``find'' partons of type $a$/$b$ with longitudinal momentum fractions $x_1$/$x_2$ inside the hadrons $1$/$2$. These probabilities cannot be calculated from first principles but can be determined from experiment. 
In the perturbative part, which can be completely calculated from first principles in perturbation theory, the hard (or partonic) scattering cross section $\hat{\sigma}_{ab \rightarrow X_{\text{p}}}\!(\hat{s};\{a,b,X_{\text{p}}\};\mu_F)$ describes the probability of two incoming partons of type $a$ and $b$ with a partonic CME (or hard scale) $\sqrt{\hat{s}}=\sqrt{x_1x_2s}$ to yield a partonic final state configuration $X_{\text{p}}$.

Due to the property of confinement, quarks and gluons cannot be observed separately in nature and exclusive hadronic final states $X_{\text{h}}$ have to be simulated: A parton shower evolves an inclusive partonic cross section into an exclusive partonic final state by simulating a cascade of soft/collinear QCD radiation, through consecutive soft and collinear parton splittings, in which each splitting product receives consecutively less energy. The parton shower evolution is determined perturbatively. At a certain scale $\Lambda_{QCD}\ll\sqrt{\hat{s}}$, approximately at the order of $\Lambda_{QCD}\sim\mathcal{O}(1\text{GeV})$, the partons in the parton shower hadronize, and may subsequently decay. The hadronization process cannot be determined from first principles, but needs to be modeled. 
In addition to the primary interaction, the hadron remnants, after extraction of the partons $a$ and $b$, can undergo further soft scatterings, so called multiple parton interactions (MPI), which form part of the accompanying underlying event and need to be modeled as well. 

In order not to be overly sensitive to the details of such models, which usually dominate the theory uncertainties, we depend on observables that are insensitive to physics in the soft/collinear limit. Such infrared-safe observables do not change in the limit in which additional soft or collinear particles are imposed. 
An important class of infrared-safe observables is thereby the class of jet observables, or simply jets. Jets can be imagined as collimated bunches of soft and collinear partons inside a "cone" of a certain "radius" around a central parton, determined by suitably chosen cone or cluster algorithms to specify the possible partonic configurations inside the jets. Several formulations to define such jet algorithms are available, most of which are infrared safe \cite{Weinzierl:2010cw}.

In the remainder of the article we will discuss the calculational concepts needed for the computation of the perturbative contributions to multi-jet production: Fixed order calculations, needed to describe the hard scattering, will be discussed in section \ref{sec:fixedorder}. The physics behind parton showers and how to interface them to fixed order calculations will be sketched in sections \ref{sec:showers} and \ref{sec:packagesNinterfaces}. The non-perturbative contributions will not be discussed any further.

\section{Fixed order calculations}
\label{sec:fixedorder}

The hard scattering for a certain partonic (sub-)process is determined predominantly by the corresponding hard matrix element, which describes the probability for the transition between the corresponding partonic initial and final state and can be calculated from first principles in perturbation theory in the strong (running) coupling constant $\alpha_s(Q)$, which exhibits small values at large scales $Q$. 
The corresponding fixed order calculation is evaluated at a certain (fixed) hard scale, and in the following, of which excellent reviews can be found in 
\cite{Weinzierl:2005dd,Weinzierl:2007vk}, 
we will thus suppress the scale argument in the strong coupling constant. An (infrared-safe) observable $O$ for $n$ partons in the final state, for example, is determined as follows:
\begin{equation}
\langle O \rangle \propto \sum\limits_{a,b} \! \int \!\! dx_1 dx_2 \; f_a(x_1) f_b(x_2) \,
\sum\limits_n \! \int\!\!d\phi_n \; O(p_1,...,p_n) \; |\mathcal{A}_{n+2}|^2 \;\;,
\end{equation}
where $|\mathcal{A}_{n+2}|^2$ denotes the hard matrix element for the scattering of $n+2$ partons ($2\rightarrow n$), which is computed in turn from the complex valued hard scattering amplitude $\mathcal{A}_{n+2}$. Summation and phase-space integration over all possible final states with $n$ partons is implied. For an observable whose leading order (LO) prediction is given by an $n$-parton tree-level amplitude $\scriptA_n^{(0)}$, the following expansions, up to relative order $\mathcal{O}(\alpha_s)$, are relevant for the calculations of the next-to-leading order (NLO) prediction:
\begin{equation}
|\scriptA_n|^2 = |\scriptA_n^{(0)}|^2 + \alpha_s2Re(\scriptA_n^{(0)\ast}\scriptA_n^{(1)})
\hspace{0.45cm}
\text{and}
\hspace{0.5cm}
|\scriptA_{n+1}|^2 = \alpha_s|\scriptA_{n+1}^{(0)}|^2 \;\;,
\end{equation}
where we have omitted the overall LO factor $\alpha_s^{n-2}$. The associated NLO contributions originate then from two different regions with respect to the phase space: One (the virtual) comes from the interference term of the virtual one-loop amplitude $\scriptA_n^{(1)}$ with the corresponding tree-level matrix element with $n$ partons, the other (the real) from the squared $(n\!+\!1)$-parton tree-level amplitude, with one additional parton in the final state, compared to the LO contribution. Note that the two NLO contributions are defined on phase spaces of different dimensionality. We can thus sum up the computation of the NLO contribution to an observable $O$ in two terms as
\begin{equation}
\langle O \rangle^{NLO} = 
\int\limits_{n} O_{n} V + \int\limits_{n+1} O_{n+1} R \;\;,
\label{eq:VR}
\end{equation}
where we have used a very condensed notation, in which $\int_n\equiv\int\!d\phi_n$ and where $V$ and $R$ denote the virtual and real contributions (encoding the corresponding observables and matrix elements) respectively. The amplitudes in $V$ and $R$ are traditionally computed from all possible corresponding Feynman diagrams, of which 
Fig.2 
shows two examples. Unfortunately the two terms in Eq.\ref{eq:VR} are ill defined if taken separately, since each contains singularities for certain regions in phase space: In the virtual part the integration over the loop momentum may lead to collinear, soft or ultraviolet singularities, e.g. if $k_i||k_j$, $|k_i|\rightarrow0$ or $|k|\rightarrow\infty$ respectively in the example in 
Fig.2. 
In the real part the integration over the additional real emission may lead to collinear or soft singularities, e.g. if $p_3||p_5$ or $|p_5|\rightarrow0$ respectively in the example in 
Fig.2. 
In an analytic integration using e.g. $D$-dimensional regularization, 
with $D=4-2\epsilon$ and $|\epsilon|\!\ll\!1$, 
these singularities manifest themselves as poles in $1/\epsilon$. 
The ultraviolet singularities in the virtual part are usually taken care of by a suitably chosen ultraviolet counterterm $V_{CT}$, such that $V$ is actually ultraviolet-finite. 
In addition, for infrared-safe observables it is known that the soft/collinear singular terms cancel between the real and the virtual contributions after integration, rendering their sum eventually finite \cite{Bloch:1937pw,Kinoshita:1962ur,Lee:1964is}. 

\begin{figure}
\centering
\includegraphics[scale=0.8]{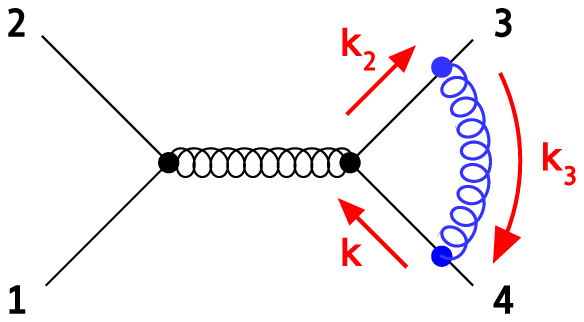}
\hspace{0.45cm}
\includegraphics[scale=1.1]{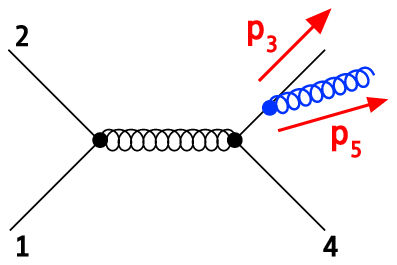}
\hspace{0.5cm}
\begin{minipage}[b]{5.5cm}
\caption{\small\it Example diagrams for virtual (left) and real (right) corrections. The degrees of freedom of the virtual as well as of the external partons are denoted by the loop four-momenta $k_i$ and the external four-momenta $p_i$ respectively.}
\end{minipage}
\label{fig:virtrealdiags}
\end{figure}

The calculation of state-of-the art processes, i.e. processes with many partons, cannot be performed analytically anymore and we need numerical methods to perform the phase-space integrals. The method of choice is hereby Monte Carlo integration \cite{Weinzierl:2000wd,Lepage:1977sw}. 
It is, however, not possible to na\"{\i}vely integrate the sum of the virtual and real contributions, since the corresponding integrands live on phase spaces of different dimensionality. The cancellation of the singular terms is thus not as trivial. A standard method to deal with such a problem is the subtraction method: 
We can rewrite the NLO observable to read
\begin{equation}
\langle O \rangle^{NLO} = \int\limits_{n+1} \big(O_{n+1}R - O_{n}A\big) +  \int\limits_{n}\big(O_{n}V + O_{n}\!\!\int\limits_{+1}\!\!A\big) \;\;,
\end{equation}
where $A$ is a simple function that has the same point-wise singular behavior as $R$. Since $O_n$ is infrared safe, i.e. $O_{n+1}\rightarrow O_n$ in the soft/collinear limit, $O_nA$ acts as local counterterm to $O_{n+1}R$ in the singular regions, rendering the first bracket finite. $A$ is thereby chosen such that it can be integrated analytically over the one-parton subspace, leading to poles in $1/\epsilon$, which then cancel against the $\epsilon$-poles from $V$, rendering the second bracket finite. Both brackets can then be separately integrated over their respective phase spaces. The singular parts of the subtraction terms are universal. The finite parts, however, can be chosen freely, which results in various variants of the subtraction method, of which several automated implementations are available. 
Relevant references can be found in  \cite{Catani:1996vz,Phaf:2001gc,Catani:2002hc,Goetz:2012uz,Frixione:1995ms,Somogyi:2005xz,Frixione:2011kh,Czakon:2009ss,Kosower:1997zr,Daleo:2006xa,Weinzierl:2005dd,Gleisberg:2007md,Frederix:2008hu,Frederix:2009yq,Hasegawa:2009tx,Frederix:2010cj,Platzer:2011bc}.

Despite the subtraction method, the real and virtual contributions still need to be constructed and especially the results of the loop integration in the virtual contribution has to be known. To compute the virtual contribution basically three methods and variants thereof exist: Traditional tensor reduction, cut-based methods and the numerical method.

In traditional tensor reduction one uses the fact that tensor integrals of the form 
\begin{equation}
I_{n,a}(\{p_i\},\{m_i\})^{\alpha_1...\alpha_{2a}} = \int d^Dk \frac{k^{\alpha_1}...k^{\alpha_{2a}}}{\prod\limits_{i=1}^{n}\big(k_i^2-m_i^2+i\delta\big)^n} \;\;, \hspace{0.25cm} \text{with} \hspace{0.15cm} D=4-2\epsilon \;\;,
\end{equation}
which result from all possible Feynman diagrams to a certain amplitude, can always be reduced to a linear combination of a certain set of known scalar integrals. The price to pay is the introduction of Gram determinants upon the determination of the corresponding coefficients. For example, reducing a three-point tensor integral of rank $2$, 
leads to the introduction of an inverse Gram determinant $|G|^{-1} \propto (p_1^2p_2^2 - (p1.p2)^2)^{-1}$, which tends to large values whenever $p_1||p_2$ and thus leads to a poor numerical behavior. However, several solutions to this type of problem exist, based on different reduction algorithms, expansion around small Gram determinants in critical regions, etc., of which several implementations are available. 
Relevant references can be found in  \cite{Passarino:1978jh,Denner:2002ii,Giele:2004iy,delAguila:2004nf,Binoth:2005ff,vanHameren:2005ed,Denner:2005nn,Weinzierl:2006qs,Diakonidis:2009fx,Heinrich:2010ax,Reiter:2010md,Fleischer:2010sq,Ossola:2007ax,Binoth:2008uq,Guillet:2013msa,Denner:2014gla,Fleischer:2011zz}.

In the cut-based methods one uses the fact that the entire amplitude can directly be decomposed into a basis of known scalar integrals. The coefficients are then obtained from tree-like objects by solving a linear system of equations numerically. These tree-like objects are universally obtained by cutting loop diagrams. For example, for a massless amplitude one needs only the coefficients in front of bubble, triangle and box integrals:
\begin{equation}
A_n^{(1)} = \sum\limits_{i,j}c_{i,j}I_2^{(ij)} + \sum\limits_{i,j,k}c_{ijk}I_3^{(ijk)} + \sum\limits_{i,j,k,l}c_{ijkl}I_4^{(ijkl)} + R_n \;\;,
\end{equation}
where the box coefficients are then for example obtained from quadruple cuts. After the box contributions have been subtracted one gets the triangle coefficients from triple cuts, and so on. Several approaches and various implementations hereof exist. 
Relevant references can be found in  \cite{Bern:1996je,Ellis:2011cr,Bern:1994cg,Ossola:2006us,Anastasiou:2006gt,Britto:2006fc,Kilgore:2007qr,Ellis:2007br,Forde:2007mi,Ossola:2008xq,Berger:2008sj,Giele:2008ve,Ellis:2008ir,vanHameren:2009dr,Badger:2011zv,Peraro:2014cba,Bern:2013pya,Badger:2010nx,Badger:2012pg}.

In the numerical method one extends the idea of the subtraction method to the virtual part by introducing local counterterms to the loop integrand in the virtual contribution as well. With $V=\int_{loop}\!V_{bare}+V_{CT}$ we can write
\begin{equation}
\langle O \rangle^{NLO} = \underbrace{\int\limits_{n+1} \big(O_{n+1}R - O_{n}A\big)}_{\langle O \rangle^{NLO}_{real}} 
+ \underbrace{\int\limits_{n,loop} O_{n}\big(V_{bare} - {L}\big)}_{\langle O \rangle^{NLO}_{virtual}}
+ \underbrace{\int\limits_{n} O_{n}\big(V_{CT} + \int\limits_{loop}\!\!L + \int\limits_{+1}\!\!A\big)}_{\langle O \rangle^{NLO}_{insertion}} \;\;,
\end{equation}
where $L$ is a simple function that has the same point-wise singular behavior as the fully singular loop integrand $V_{bare}$ in the ultraviolet, soft and collinear regions. $L$ is thereby chosen such that it can be integrated analytically over the loop-momentum space, leading to poles in $1/\epsilon$: The explicit poles from $\int_{loop}\!L$ cancel against the soft/collinear $\epsilon$-poles from $\int_{+1}\!A$ and the ultraviolet $\epsilon$-poles from $V_{CT}$. The three contributions $\langle O \rangle^{NLO}_{real}$, $\langle O \rangle^{NLO}_{virtual}$ and $\langle O \rangle^{NLO}_{insertion}$ can then be separately integrated over their respective integration spaces, where the numerical loop integration and the numerical integration over the final state phase space in $\sigma^{NLO}_{virtual}$ are performed together in one combined Monte Carlo integration. The virtual subtraction method was originally formulated on a Feynman-diagrammatic level, but later extended to be applicable directly on the amplitude 
level. The actual numerical loop integration can then either be implemented for a re-parametrized form with Feynman or Schwinger parameters or in a direct approach, where the integration is over the four-dimensional loop momenta and requires an efficient method to deform the contour of the loop integration into the complex plane where necessary. 
Relevant references can be found in  \cite{Nagy:2003qn,Assadsolimani:2009cz,Assadsolimani:2010ka,Becker:2010ng,Becker:2011aa,Becker:2012aqa,Becker:2012nf,Nagy:2006xy,Gong:2008ww,Becker:2010ng,Becker:2011aa,Becker:2012aqa,Becker:2012nf,Becker:2012nk,Becker:2011vg,Goetz:2014lla,Becker:2012bi,Freitas:2012iu,Carter:2010hi,Borowka:2012yc}.

All methods to compute the various parts to an NLO observable depend eventually on the efficient construction and evaluation of the respective integrands, be it the tree-level constituents of the LO contribution or the NLO real contribution, or the one-loop constituents of the NLO virtual contribution. The standard for many years has been to rely on traditional Feynman diagrams, where all Feynman diagrams that contribute to a certain process are taken into account. This, however, may become quite tedious for processes with many partons, since in a na\"{\i}ve implementation the computational complexity grows approximately factorially with the number of partons. In addition, considering predominantly QCD processes, the algebraic manipulations due to the underlying $\mathrm{SU}(N\!=\!3)$ gauge group can become quite cumbersome for processes with many partons. Many modern approaches that aim to cut down the computational complexity are based on the attempt to calculate entire amplitudes rather than single 
Feynman diagrams. Color decomposition offers for example a way to deal more efficiently with the underlying group theory, where one uses the fact that any QCD amplitude can in general be decomposed in a linear combination of spanning vectors in the associated group space and purely kinematical coefficients. This decomposition can be done once in the beginning of the computation, where all the group theoretical factors are computed and the linear combinations, as expansions in the number $N$ of colors, are determined. During run-time it is then only necessary to compute the kinematical coefficients for each phase-space point. This method becomes particularly efficient, if parts of the kinematical coefficients during their computation may be reused, which is possible through recursive methods. In the large-$N$ approximation one even terminates the expansions in the number $N$ of colors after the leading term, which decreases the computational effort tremendously. The kinematical coefficients can further be 
decomposed into more primitive objects, which exhibit a certain fixed ordering, i.e. all diagrams that are associated to them exhibit the same planar ordering with respect to their external legs. These primitive objects are particularly well suited to be constructed through recursive methods, of which various approaches are known. For certain helicity combinations even simple closed analytic expressions exist to compute the associated amplitudes, expressed in terms of particularly suitable spinor-helicity representations. Eventually one may use the benefits of the Monte Carlo approach, by additionally performing a sampling over the external quantum numbers (such as helicities or color degrees of freedom) through the Monte Carlo sum. 
Relevant references to the aforementioned approaches, for tree-level as well as one-loop constituents, can be found in  \cite{Dinsdale:2006sq,Duhr:2006iq,Schwinn:2007ee,Giele:2009ui,Feng:2011np,Berends:1987me,Kleiss:1988ne,Parke:1986gb,Mangano:1987xk,Mangano:1988kk,Parke:1989vn,Mangano:1987kp,Kosower:1988kh,Mangano:1990by,Kanaki:2000ms,Maltoni:2002mq,Kilian:2012pz,Melia:2013bta,DelDuca:1999rs,Keppeler:2012ih,Sjodahl:2013hra,Bern:1991aq,Bern:1994fz,Weinzierl:1999yf,DelDuca:1999rs,Ellis:2008qc,Ita:2011ar,Badger:2012pf,Bern:1996je,Ellis:2011cr,Badger:2012pg,Reuschle:2013qna,Reuschle:2014qwa,Schuster:2013aya,vanHameren:2009vq,Cascioli:2011va,Actis:2012qn,Badger:2010nx,Kleiss:1985yh,Dittmaier:1998nn,Britto:2004ap,Draggiotis:1998gr,Goetz:2012uz}.

\section{Parton showers} 
\label{sec:showers}

In this section we will sketch the motivation and physics behind parton showers (exemplarily for the case of time-like splittings) which are essential for the correct simulation of realistic detector events. Excellent general reviews on the following can, however, be found in \cite{Buckley:2011ms,ellis1996:am,dissertori2003:am,Nason:2012pr}, to which we would like to point the reader for a more thorough description. 

In order to predict detector events and observables in high energy collisions realistically, fixed order calculations are not sufficient for mainly two reasons:

1) Jet cross sections describe the inclusive production of jets: The cross section for $n$-jet production e.g. describes the production of at least $n$ jets, with a corresponding $n$-parton final state phase space. However, in order to compare to the measured detector events we need to describe the exclusive final states for exactly $n$, $n\!+\!1$, etc. jets. With fixed order calculations we will not be able to realistically describe such exclusive final states, and especially not for multi-jet configurations, due to their increasing complexity (even with modern amplitude-based methods).

2) The hard scattering is always associated with a certain hard scale, at which the corresponding fixed order calculation is evaluated. This means we need to evolve from an inclusive cross section at some hard scale $Q$ to an exclusive final state at some soft scale $\mu$, at which hadronization can take place. Due to the inherent truncation of the corresponding perturbative series and the running logarithmic behavior of the strong coupling constant, however, the evolution from one scale $Q$ to another scale $\mu$ introduces spurious logarithmic enhancements of the form $\alpha_s(\mu)^nL^m$, with $L=\log(Q^2/\mu^2)$. For a sufficiently soft scale $\mu$ the logarithms $L$ may thus easily overcome the smallness of the perturbative coupling constant $\alpha_s(\mu)$ at this scale, which leads to a divergent result for $\mu^2\ll Q^2$. Physical cross sections, however, exhibit Sudakov dampening at soft scales. A realistic description needs thus to re-sum the large logarithms to all orders in perturbation theory, 
in order to properly describe the Sudakov suppression at the soft scales 
(cf. Fig.3). 

\begin{figure}
\centering
\raisebox{-0.25cm}{%
\includegraphics[scale=0.38]{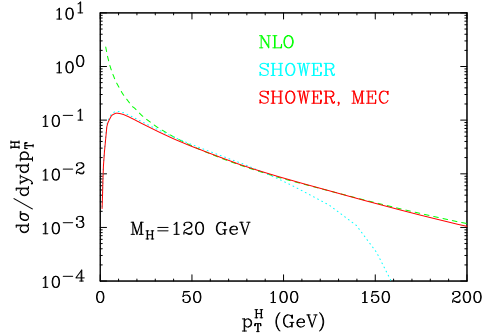}%
}
\hspace{1.4cm}
\begin{minipage}[b]{8.0cm}
\caption{\small\it The plot portrays the transverse momentum distribution of a Higgs boson in Higgs production at the LHC \cite{Nason:2012pr}, calculated at fixed NLO (green), with a parton shower (cyan), and with a matrix element corrected (MEC) parton shower (red). The NLO calculation diverges in the low $p_T$-region. The parton shower exhibits the correct Sudakov suppression in the low $p_T$-region, but fails to describe the hard region correctly. The MEC parton shower describes both regions properly.}
\end{minipage}
\label{fig:PSsudakovdamp}
\end{figure}

\begin{figure}
\centering
\raisebox{-0.1cm}{%
\includegraphics[scale=0.85]{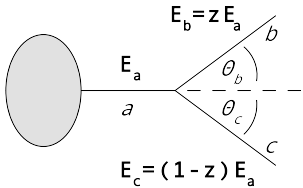}%
}
\hspace{1.5cm}
\begin{minipage}[b]{10.0cm}
\caption{\small\it A time-like splitting of a parton $a$ into two partons $b$ and $c$. For small angles: $p_a^2=t=(p_b+p_c)^2=z(1-z)E_a^2(\Theta_b+\Theta_c)^2$, where $z$ parametrizes the energy fractions in the splitting. One can then show that $d\sigma(b,c)\propto d\sigma(a)\alpha_s(t)P(z)dzdt/t$, where we note the enhancements for small values of $t$ but also for $z\rightarrow0\,\text{and/or}\,1$ encoded in the splitting function $P(z)$.}
\end{minipage}
\label{fig:splitting}
\end{figure}

A solution is presented by the parton shower approach, 
due to multiple emissions in the soft/collinear approximation: In the soft/collinear regions an $(n\!+\!1)$-parton cross section can be factorized into an $n$-parton cross section times a splitting function $P(z)$. Differentially this reads $d\sigma(\phi_{n+1})\propto d\sigma(\phi_{n})\alpha_s(t)P(z)dzdt/t$, where $t$ denotes the scale at which the splitting takes place and $z$ (respectively $1\!-\!z$) denotes the energy fraction of the splitting products, as depicted in Fig.\ref{fig:splitting}, and the phase-space element of the emission is given by $dzdt/t$ (for simplicity, here and in the following, we omit the azimuthal angle component of the emission). The splitting function $P(z)$ is enhanced for $z\rightarrow0\,\text{and/or}\,1$ (depending on the parton types of the constituents in the splitting), and in addition we note an enhancement for small values of $t$. A parton shower algorithm implements multiple emissions in the soft/collinear 
approximation, and hence the production of multi-parton final states in this approximation, through the recursive application of the above factorization, which can be sketched as follows:
\begin{equation}
PS[d\sigma(\phi_{n};t)]   
=\Delta(t,\mu)d\sigma(\phi_{n};t)+PS[\Delta(t,t')d\sigma(\phi_{n};t)\alpha_s(t')P(z')dz'dt'/t'] \;,
\label{eq:PS}
\end{equation}
where the Sudakov form factor $\Delta(t,t')$ describes the probability to evolve from a scale $t$ to a scale $t'$ without any resolvable emission, and the splitting function $P(z')$ the probability for an emission with an energy fraction $z'$. The first term in Eq.\ref{eq:PS} describes the probability to have no resolvable emission at all, between a scale $t$ and a certain cut-off scale $\mu$, whereas the second term describes the probability for at least one resolvable emission, where the integration over $z'$ and $t'$ within suitable boundaries at each recursion step is implied, which is performed through Monte Carlo integration: Choosing at each recursion step suitable values for $z'$ and $t'$ for the next emission and performing the Monte Carlo sum, where the associated weights are chosen according to the corresponding Sudakov form factors and splitting probabilities, generates eventually a corresponding exclusive final state phase space. 
A parton shower preserves therefore the value of the total inclusive cross section upon which it is applied, but redistributes the differential weights within the exclusive final state phase space, such that unitarity is preserved: $\mathcal{P}(\text{no emission})\!+\!\mathcal{P}(\text{at least one emission})\!=\!1$. Each parton has a cut-off scale $\mu\sim\Lambda_{QCD}$, below which the perturbative evolution ceases to make sense and the shower evolution is terminated. The emissions of a parton shower are naturally ordered by the values $t_n>t_{n+1}>t_{n+2}\ldots$ of the scale $t$ (also called ordering variable) at which the consecutive splittings take place, where various choices are possible (the angle between the splitting products, the transverse momentum of the splitting, etc.). Depending on those choices, a parton shower is called ``angular-ordered'', ``$p_T$-ordered'', etc. 
All parton showers exhibit the same behavior regarding the summation of the leading (large) logarithms. However, the choice of the ordering variable has an influence on how the next-to-leading logarithms are treated. 
The choice of the ordering variable will also affect how certain correlations between the splitting products are incorporated into the phase-space population. It has been shown for example, that an angular ordered shower incorporates the effects of color coherence between the splitting products correctly, but we may have small angle hard emission after large angle soft emission. A $p_T$-ordered shower on the other hand has the pleasant feature that its splitting products are ordered in ``hardness'', however the possibility to have large angle soft emission after small angle hard emission destroys the color coherence between the splitting products. A slightly different type of parton shower, the ``dipole based parton shower'', 
is based on the color-dipole picture of QCD and utilizes $2\!\rightarrow\!3$-splittings rather than $1\!\rightarrow\!2$-splittings. This type of parton shower has two advantages over the traditional approach: 
1) In $2\!\rightarrow\!3$-splittings energy-momentum conservation and the correct mass-shell conditions for each participating parton can be fulfilled simultaneously. 2) It can be ordered in hardness while at the same time implementing the correct color coherence effects. One property, however, is common to all parton shower algorithms: They utilize the large-$N$ approximation in a certain sense. However, first ideas to include certain next-to-leading terms from the expansion in the number $N$ of colors have already been formulated.

If we compare fixed order calculations with the parton shower approach we note that both exhibit necessary characteristics 
(cf. Fig.3): 
A parton shower describes many-particle final states correctly to all orders in perturbation theory, but only in the soft/collinear approximation. The fixed order approach on the other hand works very well to describe hard emission, but does not work so well in the soft/collinear regions, if we want to describe a physical result, since it fails to describe the Sudakov suppression. The obvious idea is to combine the benefits of both approaches to get a more reliable description of exclusive detector events. The general strategy is thereby to correct the first few emissions from a parton shower with the help of fixed order matrix elements (in the following shortly called matrix elements). In general, the more matrix elements are involved the better: The first few emissions are then generated with the correct weights by the matrix elements (ME), whereas the remaining multi-parton soft/collinear emissions, and hence the approximate all order summation, are described by the parton shower (PS). A na\"{\i}ve combination of ME and PS, however, leads to overlapping contributions and double counting, and one needs to take proper care of this, for which various approaches have been engineered, known under the collective terminus of ``matching'' and ``merging''.

The simplest merging algorithm consists of correcting the first emission of a parton shower by the corresponding matrix element, for example to correct the first emission of the parton shower, applied to the leading order cross section for $2$ jet production in $pp$ collisions, by the LO matrix element for $3$ jet production. The effect of such a matrix element correction (MEC) is portrayed in 
Fig.3 
for the example of Higgs production at the LHC \cite{Nason:2012pr}. 

Leading order merging consists of correcting the first few emissions of a parton shower by several LO matrix elements of various jet multiplicities. However, a na\"{\i}ve summation of the corresponding inclusive cross sections leads to double counting. LO merging prescriptions are prescriptions on how to turn inclusive cross sections exclusive before adding them.

Next-to-leading order matching consists of correcting the first emission of a parton shower by the corresponding NLO matrix element. However, both approaches contain NLO contributions, i.e. parts of the real emission contribution: Looking at the first emission of a parton shower applied to some LO matrix element and at the real emission fixed order correction to this matrix element, these contributions will overlap in the soft/collinear regions, leading to double counting. NLO matching prescriptions are prescriptions on how to construct auxiliary NLO cross sections that return the correct results without double counting upon the application of a parton shower.

The choice of the ordering variable will also affect how we have to modify a parton shower upon the various matching and merging prescriptions. Certain matching prescriptions for example depend on an ordering in ``hardness'', such that if we wanted to apply them with an angular ordered shower we would have to veto any events, produced by the shower, whose first emission is harder than a certain maximum value, which is then called a ``vetoed shower''. However, by throwing away these events we effectively constrain the phase space that is filled by the shower. This issue is solved by the concept of a ``truncated, vetoed shower''. A dipole shower on the other hand can be ordered in hardness while at the same time implementing the correct color coherence effects.

State-of-the-art nowadays are NLO multi-jet merging prescriptions, which are prescriptions on how to merge multiple NLO matrix elements and multiple LO matrix elements to a parton shower, in order to get the highest possible accuracy, of which various variants exist. 

An extensive list of relevant references to the methods described in this section can be found in 
\cite{Gribov:1972ri,Altarelli:1977zs,Dokshitzer:1977sg,Nagy:2006kb,Giele:2007di,Dinsdale:2007mf,Schumann:2007mg,Platzer:2009jq,Platzer:2011bc,Kilian:2011ka,Alioli:2012fc,Seymour:1994df,Andre:1997vh,Catani:2001cc,Lonnblad:2001iq,Frixione:2002ik,Krauss:2002up,Nason:2004rx,Nagy:2005aa,Frixione:2007vw,Lavesson:2008ah,Hoche:2010av,Hoche:2010kg,Siegert:2010mk,Hoche:2010pf,Hamilton:2010wh,Hoeche:2011fd,Lonnblad:2011xx,Platzer:2012bs,Lonnblad:2012ix,Lonnblad:2012ng,Frederix:2012ps,Hoeche:2012yf,Platzer:2013tla,Rubin:2010xp,Nagy:2012bt,Platzer:2012hp,Platzer:2012np,Platzer:2013fha}. 
References to a selection of general purpose event generators, which implement these methods can be found in  \cite{Sjostrand:2006za,Bahr:2008pv,Bellm:2013lba,Gleisberg:2008ta,Sjostrand:2014zea}.

\section{Interfacing Monte Carlo programs} 
\label{sec:packagesNinterfaces}

Recent developments in the automatization of fixed NLO calculations as well as NLO multi-jet merging make it of course necessary to have an interface to share the information between the two main ingredients. It is thereby assumed that the event generator (called Monte Carlo program in this context), which contains the parton shower algorithm, sets up the framework for a certain process and asks the fixed order program (called one-loop provider in this context) for the NLO matrix element during run-time. For this purpose the BLHA standard has been created, and recently extended \cite{Binoth:2010xt,Alioli:2013nda}. The advantage of having an automated standard is that the computation of the whole process chain is less error prone. The Monte Carlo program (MC) steers the setup of all necessary hard matrix elements, and provides for the numerical phase-space integration and the parton shower (if needed). The one-loop provider (OLP) provides the hard LO as well as NLO matrix elements on request for a certain 
phase-space point during the numerical integration. The initial setup and the list of required matrix elements is communicated once at the beginning of the computation, via a simple order and contract file system, the results of the individual matrix elements per phase-space point via external function calls during run-time. 
Besides the general purpose event generators, of which we have mentioned a selection in the previous section, there exist various Monte Carlo matrix element generators, utilizing the various fixed order approaches described in section \ref{sec:fixedorder}, a selection of which, used for tree-level as well as one-loop calculations, can be found in  \cite{Arnold:2008rz,Baglio:2014uba,Arnold:2011wj,Kilian:2007gr,Mangano:2002ea,Krauss:2001iv,Gleisberg:2008fv,Cullen:2014yla,Alwall:2014hca,Campbell:2002tg,Alioli:2010xd,Cascioli:2011va,Actis:2012qn,Badger:2012pg,Badger:2012pf}. Noteworthy hereby are those NLO programs, which may serve as independent OLP programs or come with their own parton shower capabilities \cite{Arnold:2008rz,Baglio:2014uba,Arnold:2011wj,Kilian:2007gr,Cullen:2014yla,Alwall:2014hca,Bern:2013pya,Cascioli:2011va,Actis:2012qn,Badger:2012pg,Badger:2012pf}. 
A selection of processes, which have been computed in a combined effort between MC and OLP programs can be found in  \cite{Butterworth:2014efa,Bern:2013gka,Badger:2013yda,Bern:2011ep,Hoeche:2014qda,Cullen:2013saa,Berger:2010zx,Campanario:2013fsa}.

\section{Conclusion}

It is probably fair to say that fixed NLO QCD calculations are now in a state in which LO calculations have been about a decade ago. Various important processes, which have been stated in an NLO ``wish list'' \cite{Buttar:2006zd}, have been successfully calculated during this ``NLO revolution'', effectively closing said wish list. New important processes have been agreed on in a new NLO ``wish list'' \cite{Butterworth:2014efa}, with a new focus on mixed QCD and electroweak corrections but also on NNLO QCD corrections. Accordingly, methods to combine the benefits of fixed NLO QCD calculations with existing parton shower implementations have been developed and interface standards have been formulated. Several studies, using those interface standards, have been performed already. The full potential with which the new standards present us, however, has yet to be tapped into, i.e. with these new possibilities at hand we may now focus all the more on yet to be answered issues of rather physics related questions, 
as for example on how to assign scale uncertainties in multi-scale problems, etc., and not so much on the technical details behind the underlying calculations anymore.

\section*{Acknowledgments}

C.R. would like to thank the committees of the ACAT 2014 workshop, as well as the convenors for track $3$, for the opportunity to present this overview. The work of C.R. is supported in part by the BMBF.

\section*{References}

\bibliography{ACAT2014_CR}

\end{document}